\documentclass[pra]{revtex4}

\usepackage{graphicx}
\usepackage{amsmath}
\usepackage{epsfig}
\usepackage{dcolumn}  
\usepackage{bm}       
\usepackage{color}
\usepackage{bbold}
\usepackage{epstopdf}
\usepackage{color}

\newcommand{\ket}[1]{{\left\vert {#1} \right\rangle}}

\begin{document}

\title{Toward quantum simulations of biological information flow}

\author{Ross~Dorner$^{1,2}$, John~Goold$^{2,3}$ and Vlatko~Vedral$^{2,4}$}

\affiliation{$^1$Blackett Laboratory, Imperial College London, United Kingdom,\\
$^2$Clarendon Laboratory, University of Oxford, United Kingdom,\\
$^3$ Physics Department, University College Cork, Cork, Ireland,\\
$^4$ Centre for Quantum Technology, National University of Singapore, Singapore}

\date{\today}

\begin{abstract}
Recent advances in the spectroscopy of biomolecules have highlighted the possibility of quantum coherence playing an active role in biological energy transport. The revelation that quantum coherence can survive in the hot and wet environment of biology has generated a lively debate across both the physics and biology communities. In particular, it remains unclear to what extent non-trivial quantum effects are utilised in biology and what advantage, if any, they afford. We propose an analogue quantum simulator, based on currently available techniques in
ultra-cold atom physics, to study a model of energy and electron transport based on
the Holstein Hamiltonian
By simulating the salient aspects of a biological system in a tunable laboratory setup, we hope to gain insight into the validity of several theoretical models of biological quantum transport in a variety of relevant parameter regimes.
\end{abstract}
\maketitle 

\section{Introduction}
\label{sec:intro}

The observation of long-lived quantum coherence within the photosynthetic Fenna-Matthews-Olson complex \cite{Engel:07}, even at room temperature \cite{Panitchayangkoon:10}, has raised the possibility of quantum coherence playing a functioning role in the transport of energy within biological molecules. This runs contrary to traditional semi-classical, diffusive models of biological energy and electron transport  based on purely stochastic dynamics \cite{Forster:48, Marcus:85, Hopfield:74}, spurring the development of new microscopic models that allow for quantum mechanical effects.

\begin{figure}[tb]
\includegraphics[scale=0.4]{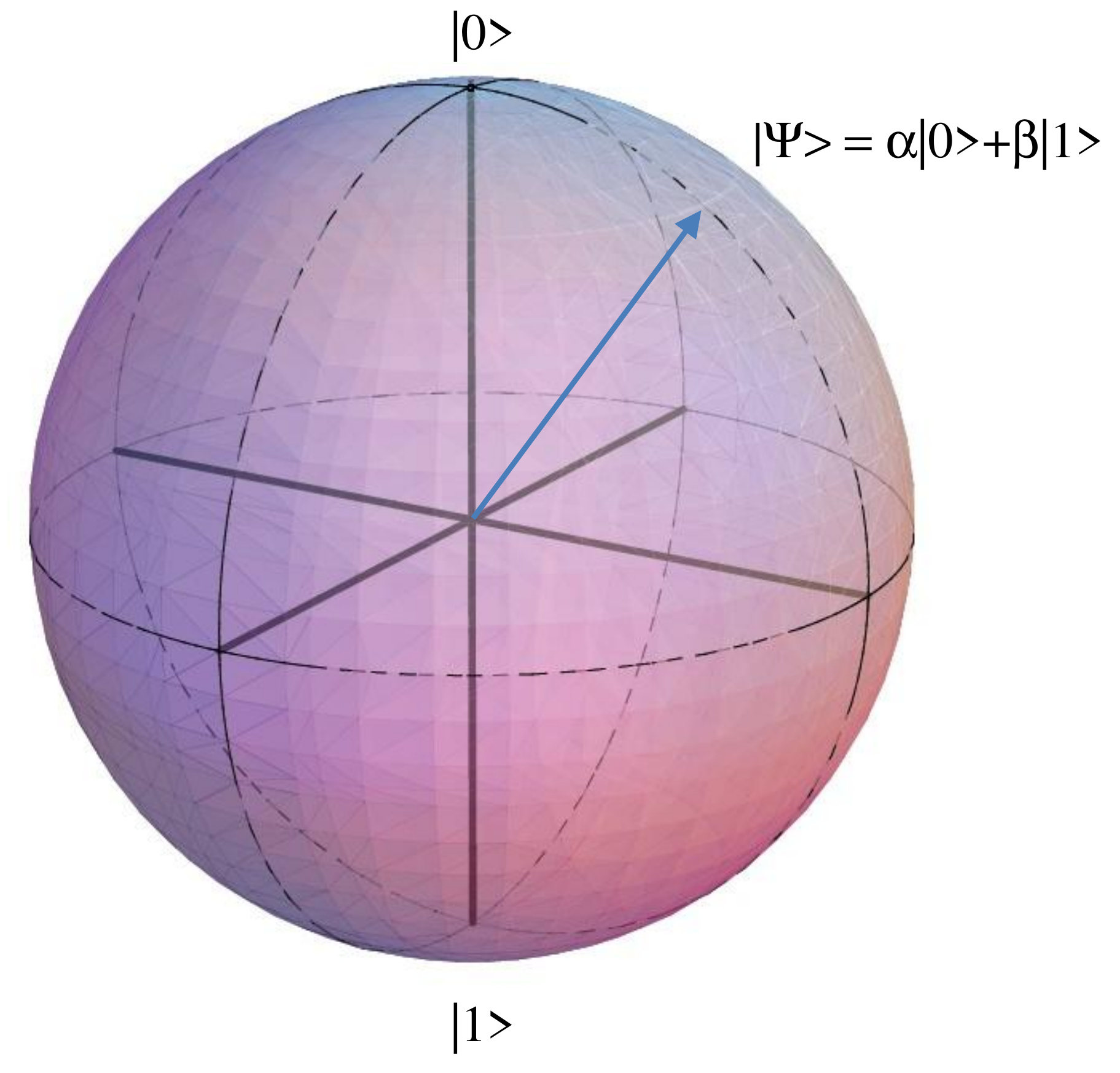}
\caption{The Bloch sphere provides a geometric representation of a qubit. In classical information theory a bit can exist in one of two possible states `0' or `1', represented by the poles of the sphere. The state of a qubit can be any one of an infinite number of points along the surface of the Bloch sphere. The poles of the sphere represent the computational states $\ket{0}$ and $\ket{1}$ of the qubit and all other points (such as that indicated by the blue arrow) describe a coherent superposition of these states $\ket{\psi}=\alpha\ket{0}+\beta\ket{1}$, subject to the normalisation condition $|\alpha|^2+|\beta|^2=1$. The counter-intuitive property of quantum coherence leads to many uniquely quantum phenomena, such as the superior power of quantum information processing protocols or the formation of Bose-Einstein condensation in gasses of ultra-cold atoms.}
\label{fig:simfig1}
\end{figure}

Coherence in quantum mechanics is a direct manifestation of the superposition principle, a defining feature of quantum mechanics that allows quantum particles to exist in several different states simultaneously. This lies in stark contrast to our `classical' perception, where the objective properties of particles have well defined values.
For instance, in classical information theory, information is encoded within a {\it bit} that exists in one of two computational states, `0' or `1', at all times. Following classical intuition,  the act of measurement deterministically reveals the well defined property of the bit value. This logic fails when the states of a two level quantum object, such as an electron spin, are used to encode information. This quantum bit or {\it`qubit'} may exist in either of the two computational states, $\ket{0}$ or $\ket{1}$, or any of an infinite number of superposition states comprising a linear combination of them (see Fig. \ref{fig:simfig1}). Measurement of the qubit state `collapses' the superposition to either one of the computational states with a certain probability. Despite the loss of information upon collapse, the ability of the qubit to exist in a superposition of both $\ket{0}$ and $\ket{1}$ gives quantum computation an inherent parallelism over analogous computations using only classical bits \cite{Deutsch:85}. Accordingly, the field quantum information theory has exploited quantum coherence to develop information processing schemes of superior power to their classical counterparts \cite{Nielsen:00}.

Quantum coherence is also the driving force behind many physical phenomena such as high temperature superconductivity in certain synthetic materials, Bose-Einstein condensation in clouds of ultra-cold atoms and various transport processes in condensed matter systems. In the particular case of transport, the presence of quantum coherence can lead to a considerable increase in the speed of the process. This is illustrated by the one-dimensional random walk, a model for the transport of a single particle. In a classical random walk, a `walker' (representing a particle) starts at the origin of an infinite line. For simplicity, the line is divided into discrete points, as is time. At each point in time a two-sided coin (a bit) is flipped and the walker steps one unit to the left upon obtaining a `heads' outcome, or one unit to the right for `tails'. The coin is flipped a further $M-1$ times, following the same rules, and the position of the walker is measured after $M$ steps. Repeating this experiment many times yields a probability distribution for the position of the walker that is binomial with a width proportional to $\sqrt{M}$, characterising diffusive transport.

The quantum version of a random walk \cite{Kendon:07}, differs in the key respect that the quantum coin (qubit) may exist in a coherent superposition of `heads' and `tails', dictating that, in general, the walker must step in both directions simultaneously at each point in time. After $M$ steps the position of the walker is thus a coherent superposition of all positions within $M$ points of the origin. Measurement of quantum walker's position collapses the superposition to one well defined location and the resulting probability distribution  is obtained by repeating the experiment many times. In this case, the width of the probability distribution is proportional to $M$, i.e., a quantum walk exhibits ballistic transport, offering a $\sqrt{M}$ `speed-up' due to the presence of quantum coherence. In a realistic transport process, such as the transfer of electrons in a metal, the coupling of the electron to an `environment' of quantised vibrational modes (phonons) leads to `decoherence' of the superposition across many locations. This arises from the phonons performing `weak' measurements on the position of the electron and tends to reduce quantum coherent superposition states to classical statistical mixtures. Accordingly, decoherence leads to a crossover from ballistic transport to classical diffusive behaviour with a timescale that depends strongly on the microscopic details of the material in question.

Although purely ballistic transport in biological molecules at room temperatures is improbable, it is possible that some biological transport processes operate between the purely classical diffusive regime and the quantum ballistic regime, at least partially exploiting the presence of quantum coherence. Models along these lines have been proposed to study excitonic energy transport in photosynthetic complexes \cite{Chin:2010, Caruso:2010, Asadian:2010, Rebentrost:2009, Giorda:11} and electron transport in metalloproteins \cite{Dorner:11}. Further studies have investigated the role of quantum entanglement in these settings \cite{Caruso2, Sarovar:2010}.

Remarkable as the experimental observation of quantum coherence in a room temperature biological system may be \cite{Panitchayangkoon:10}, there is still little empirical information regarding the role of quantum coherence in this context or the precise mechanism of transport. This is partly due to the complexity of biological systems, making it difficult to isolate distinct parts of the system in order to ascertain the dominant contributions to the transport. A further interesting question is how widespread quantum effects in biology might be. Is it possible that similar effects to those observed in photosynthetic energy transport are also exhibited in some biological electron transport processes?
An alternative means of investigating quantum effects in biological systems is the use of a quantum simulator. Following the original idea of Feynman \cite{feynmann}, a quantum simulator describes the construction of a relatively simple and experimentally controllable quantum system that encapsulates the essential features of the system of interest. This raises the prospect of mimicking the key features of biological energy and electron transport in an experimentally tunable setup, allowing the investigation of a wide range of relevant parameter regimes to gain insight into the dominant contributions towards transport. This extends from the current use of classical simulation which yields reliable results under only some of the conditions that are relevant to biology (see Sec. III).

Ultra-cold quantum gases have emerged as ideal an candidate for designing controllable experiments to simulate effects
in condensed matter physics \cite{bloch}. Drawing upon a range of recently developed techniques, the parameters of certain Hamiltonians
can be tuned with an unprecedented precision, allowing for the exploration of phase 
diagrams synonymous with condensed matter systems. A celebrated example in this direction was the observation of 
the superfluid-Mott insulator transition using cold atoms trapped in an optical lattice \cite{Greiner:02} (see Sec \ref{sec:therm}).
In addition to the experimental control afforded by ultra-cold atomic gases, the high degree of isolation of the system from its environment prevents the onset of unwanted decoherence and allows for excellent time resolution of quantum dynamics.

In this article we propose a quantum simulator, based on techniques currently employed in the field of experimental ultra-cold atomic physics, to simulate quantum coherent energy and electron transport based on the Holstein Hamiltonian 
\cite{Holstein:59}. This model is used as a prototype for excitonic energy transport in photosynthetic complexes \cite{Chin:2010, Caruso:2010, Asadian:2010, Rebentrost:2009, Giorda:11} and electron transport in metalloproteins \cite{Dorner:11}.
The rest of this article is structured as follows: In Sec.~\ref{sec:transport} the Holstein Hamiltonian is introduced as a description of biological energy and electron transport. Sec.~\ref{sec:simu} introduces the basic ideas of quantum simulation before, in Sec.~\ref{sec:therm} we explore the possibility of a quantum simulator of biological energy and electron transport based on current technology in ultra-cold atomic physics.

\section{Transport mechanisms in biomolecules}
\label{sec:transport}

\begin{figure}[tb]
\includegraphics[scale=0.7]{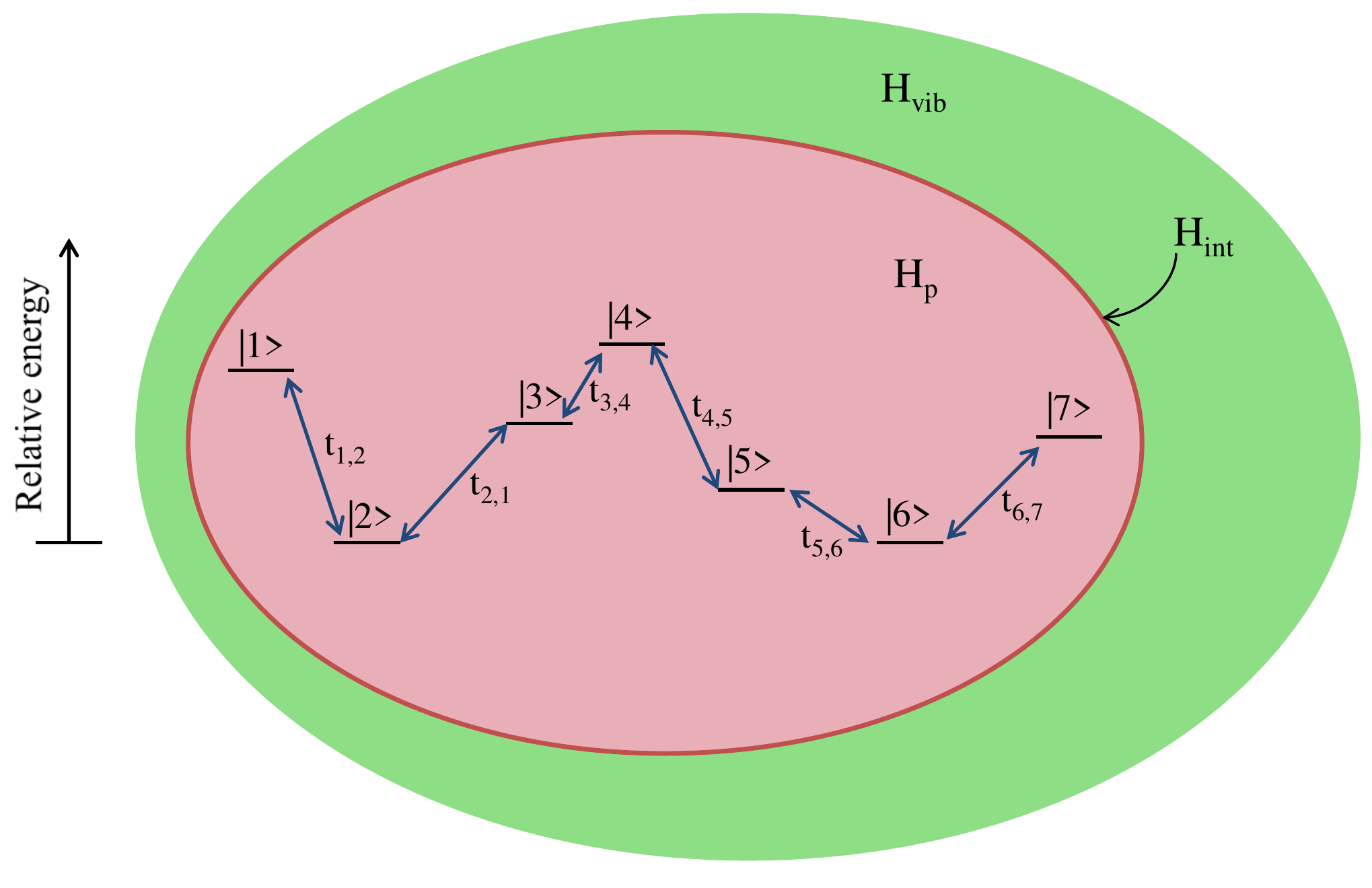}
\caption{A schematic view of a particular instance of the Holstein Hamiltonian (Eq. \eqref{eq:hampart}) describing biological energy and electron transport. Here, the system (inner, pink region) constitutes a disordered linear seven-site tight-binding chain described by $H_\textrm{p}$. The sites are labelled $\ket{i}$, $1\leq i \leq 7$, on-site energies vary from site to site and blue arrows represent coherent couplings between sites. The system is embedded within an environment (outer, green region) described by the Hamiltonian $H_\textrm{vib}$ and the two regions couple via the interaction term $H_\textrm{int}$.}
\label{fig:simfig2}
\end{figure}

Biomolecular energy and electron transport are both well studied for their relevance to emerging technologies, biomedicine and for their key roles in a variety of chemical processes in biology.
The overwhelming complexity of many biological processes, however, necessitates their simplification to a few key components for the construction of physical models. Following these lines, many biomolecules involved in  energy or electron transport
conform to a general gross structure constituting a `system' of sites embedded within an `environment' of protein and surrounding solvent (see Fig.~\ref{fig:simfig2}). The study of transport is then reduced to the study 
of quantum coherent motion of a particle among the sites of the system in a manner reminiscent of a quantum random walk. 
Crucially, the coupling of the system to its environment (containing millions of atoms) `washes out' the signature of quantum coherence by introducing non-unitary dynamics to the system. These manifest themselves as dephasing (coherence destroying) and incoherent tunnelling (stochastic hopping) processes.
Studies accounting for these effects have lead to the notion of environmentally assisted transport \cite{Chin:2010, Caruso:2010, Asadian:2010, Rebentrost:2009, Giorda:11, Paternostro:2011, Dorner:11}, whereby the interplay of coherent transport and coherence destroying processes lead to enhanced transfer 
rates compared to those predicted with semi-classical or purely coherent models.

The association of quantum effects with an enhancement of transport rate has drawn analogy with the speed-up offered by quantum computers over their classical counterparts \cite{Sension:07}. However, in biological energy or electron 
transport it is the {\it interplay} of quantum coherence and classical noise, traditionally considered deleterious to phenomena utilising quantum mechanics, that leads to the greatest speed-up. This implies that the coupling  between the system and environment determines to a large extent the reduced dynamics of the system. 

We use the Holstein Hamiltonian as a model that is general enough to describe both excitonic energy transfer in photosynthetic complexes and electron transport in metalloproteins, a class of biomolecules containing chains of metallic redox centres. Metalloproteins have several functions in biology, including respiratory complex I, an enzyme that catalyzes a crucial step in respiration in mitochondria and many aerobic bacteria \cite{Sazanov:06}. Understanding electron transport in metalloproteins is beneficial to the engineering of molecular electronic devices and, due to their key role in a multitude of important biological processes, may provide insight into the factors that contribute to the onset of various diseases \cite{Lin:2006}. This model can be used to study general features of the time evolution of these systems. 
 
The so-called interacting spin-boson model is well-studied as a model of energy and electron transfer in condensed phases, see for example \cite{Nitzan}. The Holstein Hamiltonian is a straight forward extension of the spin-boson Hamiltonian to systems of more than two sites. The Holstein Hamiltonian can be partitioned into particle, environmental and interaction contributions as
\begin{equation}
H=H_\textrm{p}+H_\textrm{vib}+H_\textrm{int}.
\label{eq:hampart}
\end{equation}
The particle term of the Holstein Hamiltonian $H_\textrm{p}$ is the well-known disordered tight-binding Hamiltonian for a network of $N$ sites,
\begin{equation}
H_\textrm{p}=\sum_{i}^N E_i \frac{\left( \mathbb{1}_i- \sigma_i^z \right)}{2}+h\sum_{i \neq j}^{N}  t_{i,j} \left( \sigma^+_i \sigma^-_{j} +\sigma^+_{j} \sigma^-_{i} \right).
\label{eq:hamelec}
\end{equation}
Where $\sigma^i$, $i\in \{x,y,z\}$ are the Pauli matrices acting at the $i^{th}$ site and $\sigma_i^\pm = \frac{1}{2}(\sigma_i^x \mp \sigma_i^y)$, $h$ is Planck's constant, $E_i$ refers to the on-site energy of a particle localised at the $i^{th}$ site and $t_{i,j}$ is the amplitude of coherent coupling of the particle between the $i^{th}$ and $j^{th}$ sites. 
The parameters $t_{i,j}$ and $E_j$ can already contain a certain amount of complexity that accounts for the vibrational structure of each site \cite{Dorner:11}. Eq. \eqref{eq:hamelec} thus describes the coherent motion of a quantum particle through an arbitrary network of sites. In the case of photosynthetic energy transfer it describes the motion of an exciton through a two-dimensional network of seven chromophores, while for respiratory complex I it describes the tunnelling of an electron through a linear chain of seven redox centers. The assumption that photosynthetic energy transfer operates in the single excitation manifold and its relevance to quantum effects in such models has been critiqued in Ref. \cite{Tiersch:2011}.

The environment is typically modelled as an infinite set of harmonic oscillators, each having characteristic frequencies $\nu_{k}$ and corresponding bosonic creation and annihilation operators, $a^\dagger_{k}$ and $a_{k}$, for the $k^{th}$ mode.
The environmental component of the Holstein Hamiltonian is then
\begin{equation}
H_\textrm{vib}=h\sum_{k} \nu_{k}  a^\dagger_{k} a_{k}.
\label{eq:hamvibr}
\end{equation}
Finally, the interaction part $H_\textrm{int}$ of the Holstein Hamiltonian accounts for the system-environment coupling
\begin{equation}
H_\textrm{int}=
\sum_{i}^N\sum_{k} g_{i,k} \frac{\left( \mathbb{1}_i- \sigma_i^z \right)}{2} \left(a_{k} + a^\dagger_{k} \right).
\label{eq:interham}
\end{equation}
Where $g_{i,k}$ describes the coupling strength between a particle at site $i$ and the $k^{th}$ phonon mode. The effect of the environment on the particle transport is completely governed by the strength of the couplings $ g_{i,k}$ and the density of modes of the environment (the number of oscillators per unit frequency). These quantities are, in general, difficult to extract for a biological system given the large number of atoms involved and poor degree of experimental access.
Nontheless, theoretical models of the dynamics rely intricately on the precise behavior of these quantities. On this basis, the construction of quantum simulator that is able to controllably mimic each of the components of the Hamiltonian would not only serve to demonstrate the technological applicability of environmentally assisted transport but also provide a testbed for elucidating the effect of environments with differing density of modes and couplings. A quantum simulator may also provide a means of probing different parameter regimes to help ascertain how widespread quantum effects in biology could be.

\section{Quantum simulators: the state of the art}
\label{sec:simu}

Even if the Hamiltonian of a disordered quantum many-body system such as a biomolecule can be constructed, there still remains the herculean computational task of simulating the quantum state on a classical computer. The difficulty is due to the large classical computational memory resources required to store the quantum state, which scales exponentially with the size of the system. This can be seen by considering the memory required to store the quantum state of an arbitrary $n$ qubit system $\ket{\psi^n}$ \cite{kendon}.  In general, the state is described as a superposition of $2^n$ basis states $\ket{k}$, labelled by the integers $0\leq k\leq 2^{n-1}$ that span the Hilbert space of the system
\begin{equation}
\ket{\psi^n}=\sum_{k=0}^{2n-1} \alpha_k \ket{k},
\end{equation}
subject to the normalization condition $\sum_k |\alpha_k|^2=1$, $\alpha_k \in \mathbb{C}$ $\forall k$. Assuming that the real and imaginary components of each $\alpha_k$ are represented by decimal32 floating-point variables we have the following formula for the number of classical bits $B$ required to store the state of an arbitrary $n$ qubit system.
\begin{equation}
\mbox{B} = 2^{n}\times2\times32= 2^{n+6}.
\end{equation}
In contrast, the memory required to store the state of a classical system of particles scales linearly with the size of the system. A high-specification modern day desktop computer may have $16 \mbox{ Gbytes}$  ($2^{37} \mbox{ bits}$) of memory. This allows the state of a classical system containing many millions of particles to be stored, while being limited to only an $n=31$ qubit system {\it at most}. The problem is compounded when the system is coupled to a large environment which can extend the Hilbert space to infinite dimension. Furthermore, calculating the time evolution of even moderately sized quantum systems can be a computationally intensive process.
Although current theoretical and numerical approaches to open quantum systems dynamics can circumvent some of the problems associated with the classical simulation of quantum systems using variational master equations, numerical path integration and adaptive truncation of the Hilbert space (see Refs.~\cite{Ishizaki:09, McCutcheon:11, Aspuru:2011} for example), efforts are still ongoing to accurately describe open system dynamics in all relevant parameter regimes and environments in the context of biological energy and electron transport.  A possible solution to these problems was proposed 
over a quarter of a century ago by Feynman \cite{feynmann}, who suggested that a controllable laboratory device obeying the laws of {\it quantum mechanics} could be programmed to efficiently simulate the physics of another quantum system, avoiding the aforementioned pitfalls associated with classical simulation. 
Since then, spectacular advances have been made in the experimental coherent control of quantum systems. This has lead to the experimental realisation of a quantum simulator, with a plethora of candidate platforms becoming available \cite{kendon, Buluta}.

Quantum simulators fall into two categories; analog and digital. A {\it digital} simulator is a more recent refinement of Feynman's original idea that draws 
upon ideas of quantum computation to use systems of qubits to encode the quantum state of an arbitrary 
many-body system. Digital simulation relies on the ability to decompose the Hamiltonian of the system to be simulated into a series of one and two qubit operations that are executed in a stroboscopic fashion on an ensemble of laboratory qubits. The 
details of this procedure were fleshed out by Lloyd in Ref. \cite{lloyd} along with an extension to simulate an open quantum system. Very recently these ideas have been extended to the realm of dissipative systems \cite{eisert}, of direct applicability to transport in biological systems. The realisation of a universal quantum simulator capable of simulating arbitrary quantum systems, closed or open, would represent a landmark achievement. Progress in this direction has been made recently in strings of cold trapped ions \cite{lanyon}. 

{\it Analog} quantum simulators are experimental systems with a direct mapping between the state and dynamics of the many-body system to be simulated to the laboratory device. Although analog simulators are limited by their ability to simulate only a very specific class of process, they have been implemented experimentally in a number of different guises \cite{Buluta}.
The key components required for performing an analog quantum simulation are 
\begin{itemize}
\item The ability to prepare your quantum simulator in some well defined initial state.
\item The ability to accurately read out a final state after some finite time of dynamical evolution.
\item The ability to extract the desired information from the simulator. 
\end{itemize}
In particular, the extraction of the desired information is crucial as it conceivable that while a device accurately simulates the system of interest, the relevant information is inaccessible due to experimental imperfections.
For example, in the case of quantum transport the time dependent density dynamics of the particle may be extracted without obtaining information related directly to the spatial coherence of the particle. Existing proposals on how quantum simulation may be used in biological settings include simulation of the noise assisted transport in an array of coupled optical cavities \cite{MPlenio:2011} and the quantum-aided simulation of molecular dynamics for large biomolecules \cite{Harris:10}.
In the following section we propose an analog quantum simulator based on  currently available techniques in ultra-cold atomic physics that is capable of providing information on the particle dynamics and spatial coherence 
in biological energy and electron transport described by the Holstein Hamiltonian (Eq. \eqref{eq:hampart}).

\section{Ultra-cold atomic simulations of biological transport}
\label{sec:therm}

Ensembles of trapped ultra-cold atoms have emerged as prime candidates for the analog simulation of condensed matter systems \cite{ols, lewenstein}. A textbook example is the microscopic model of electricity based on electron-phonon interactions \cite{Mahan:00}. The first
component required for a faithful simulation is a periodic potential, provided by an optical lattice. An optical lattice is a spatially periodic light pattern formed by the interference of two counter propagating laser fields. When atoms are cooled to ultra-cold temperatures they are trapped at the potential minima. The resulting arrangement is reminiscent of the standard crystal lattice synonymous with solid state physics. The properties of the lattice, such as the lattice spacing and the depth, can be tuned by manipulation of the external laser fields. The interactions between the particles can also be controlled. This is possible by means of Feshbach resonances which are generated by applying an external field to the ensemble. The resonance occurs when the energy of a bound state of the inter-atomic scattering potential between atoms is equal to the kinetic energy of a colliding pair of atoms. The combination of these techniques allows the replication of the Hamiltonian described in Eq. \eqref{eq:hamelec} to describe the tunnelling of an atom among the resonant sites of the optical lattice.

The optical lattice differs from a real lattice in the key respect that it does not possess any phonon modes. To remedy this, Jaksch {\it et al.} \cite{d1,d2,d3} have proposed the use of single and multiple neutral `impurity' atoms trapped in an optical lattice and immersed in an auxiliary Bose-Einstein condensate formed from a different atomic species. Here the lattice is far detuned from the atomic transitions of the Bose-Einstein condensate atoms and the low energy excitations of the Bose-Einstein condensate form a quadratic phonon bath in the form of Eq.~\eqref{eq:hamvibr}. The phonons then couple to the impurity atoms within the optical lattice via an interaction term identical to Eq.~\eqref{eq:interham} where . The coupling parameter $g_{i,k}$ is determined by the $s$-wave scattering length of the impurity atoms with the atoms of the Bose-Einstein condensate and the overlaps of the spatial wavepacket of the impurity and the phonon modes of the Bose-Einstein condensate (for further technical details see Refs. \cite{d1,d2,d3}). This setup provides a tunable analog simulator of the Holstein Hamiltonian, allowing the study of system-environment like interactions in a controlled manner. The intra- and inter- species interactions may be tuned and even `turned off' using a combination of both the trapping geometry and the available Feshbach resonances. Dynamics of the impurities are initiated by providing an effective voltage in the form of a lattice tilt and the impurity dynamics can be tracked using known experimental techniques \cite{Palzer:09,ssdp}. By varying the dispersion relation and temperature of
the Bose–Einstein condensate, a crossover between
quantum coherent and classical diffusive transport
is observed~\cite{d1,d2,d3}.

The simulation of biological energy and electron transport introduces a number of specific difficulties to the setup described above.
Biological transport typically involves {\it mesoscopic} systems containing only a small number of sites (seven in the case of respiratory complex I). Furthermore, biological transport is marked by the presence of disorder in the particle energies and inter-site distances, this being reflected in disorder in the on-site energies $E_i$ and coherent couplings $t_{i,j}$ contained in Eq.~\eqref{eq:hamelec}.
Neither of these possibilities are afforded by standard optical lattice techniques. Recently however, a method of generating arbitrary optical potentials for ultra-cold atoms was implemented by projecting a holographic mask through a specialised imaging system \cite{ greiner_amaz1, greiner_amaz2}. Here, individual atoms in a tight-binding-like lattice were detected with almost perfect fidelity. It is also possible to make real-time simulations of the dynamics. In principle this technique could be used to generate a small lattice with a level of disorder that is typical of the parameters in the Hamiltonian (Eq.~\eqref{eq:hamelec}) describing biological transport. A single particle could be trapped in this geometry and, as described above, a phonon bath in the form of an auxiliary quantum gas can be introduced. Alternatively, rather than  than a single impurity hopping in a lattice, the Holstein Hamiltonian could be implemented with a single excitation shared across atoms trapped in each lattice site and this excitation interacting with the excitations of the auxiliary gas.

While it is important to state that such a simulator does not exactly describe {\it in situ} biological energy and electron transport, it can provide insight into several features of biological quantum transport that are currently the subject of active research.
One of the features debated is the principle of noise-assisted transport. Here it is argued that noise stemming from the protein and water environment coupled with a finite degree of particle coherence can lead to an enhanced transfer rate across the biomolecule \cite{Chin:2010, Caruso:2010, Asadian:2010, Rebentrost:2009, Giorda:11, Paternostro:2011, Dorner:11}. A related avenue of interest is the validity of various approximations in certain parameter regimes of the environment \cite{Ishizaki:09, McCutcheon:11}. Both of these concepts may be studied by utilizing the inherent controllability of systems of ultra-cold atoms and optical lattices. This allows experimental control of the on-site energies and coherent coupling terms of Eq. \eqref{eq:hamelec} in addition to allowing the dispersion relation, temperature and the impurity-phonon interaction within the environment to be tuned or even turned off, allowing manipulation of the parameters in Eqs. \eqref{eq:hamvibr}-\eqref{eq:interham}. Imaging the dynamics of the impurity atoms may allow the identification of different isolated contributions to the transport process. In addition, it may be possible to gain information on the degree of quantum coherence of the impurity by measuring its momentum distribution, and hence its first order correlation function. 

\section{Conclusion}
\label{sec:conclusion}

We have proposed an analog quantum simulator based on currently available methods in the field of ultra-cold atomic physics, to study several salient aspects of quantum coherent energy and electron transport in biology. Using an experimental system consisting of an ultra-cold impurity atom within an arbitrarily shaped optical lattice \cite{greiner_amaz1, greiner_amaz2} and submerged in a Bose-Einstein condensate, a fully tunable realisation of the Holstein Hamiltonian (Eq. \eqref{eq:hampart}) may be achieved \cite{d1,d2,d3}. Such a system can serve as a testbed for models of biological quantum transport in various parameter regimes \cite{Ishizaki:09, McCutcheon:11}. More generally, the notion of faithfully replicating {\it in situ} biological energy and electron transport using a quantum simulator remains an interesting open question. The development of a universal digital quantum simulator \cite{lloyd} would hypothetically allow the simulation of a multitude of relevant Hamiltonians, though the often overwhelming complexity of biological systems would still present significant difficulties in `programming' such a simulator to exactly reflect the corresponding biological system. In the shorter term, analog quantum simulators can be realised using current technology and may afford some valuable insight into the rapidly expanding area of research concerning quantum effects in biology.

\end{document}